\documentclass[lettersize,journal]{IEEEtran}
\usepackage{amsmath,amsfonts}
\usepackage{algorithmic}
\usepackage{algorithm}
\usepackage{array}
\usepackage{textcomp}
\usepackage{stfloats}
\usepackage{url}
\usepackage{verbatim}
\usepackage{graphicx}
\usepackage{cite}
\usepackage{subfigure}
\usepackage{caption}

\begin{document}

\title{Servers Placement Scheme Based on All-pay Auction Framework in Mobile Edge Computing}

\author{Yun Xia
}



\maketitle

\begin{abstract}
Task offloading plays a pivotal role in mobile edge computing, enabling terminal devices to enhance task execution efficiency and conserve energy. However, servers are reluctant to offer services without compensation. Currently, pricing mechanisms are commonly employed to incentivize servers to serve terminal devices, with servers earning revenue through payments from these devices. Given the rapid surge in terminal devices, determining the optimal number of servers placement for service providers (SPs) to maximize revenue is crucial. In this paper, we propose a server placement scheme based on an all-pay auction framework. Experimental simulations reveal that an optimal server-user ratio of approximately 25\% maximizes SP profits.
\end{abstract}


\section{Introduction}

\IEEEPARstart{M}{obile} edge computing (MEC) relocates cloud computing from the mobile core to the network edge, providing nearby users with computing, storage, and distribution services to minimize network delay and enhance data processing \cite{MEC1, MEC2}. Task offloading, a crucial aspect of MEC, transfers resource-intensive and delay-sensitive tasks to edge servers (ESs) to leverage their computing power and improve task efficiency. However, task offloading faces challenges: ESs may be unwilling to execute tasks without compensation, and end users (EUs) hesitate to offload if costs are high. Additionally, balancing ES supply with EU demand \cite{SD} is crucial for the service provider (SP) to maximize revenue. Too few ESs leads to unmet demand and lost revenue, while excess supply results in low bids and idle ESs with fixed costs, reducing SP profits. Determining the optimal server-user ratio is essential to maximize SP profits while maintaining market balance.
Currently, pricing incentive mechanisms are widely used, where EUs pay ESs for services to encourage task execution \cite{P1, P2, P3}. However, these schemes often involve one-on-one pricing between EU and ES, neglecting the overall revenue. Instead, the SP can offer multiple ESs to serve multiple EUs, focusing on the total revenue of all ESs rather than individual ES revenue.

In this paper, an all-pay auction-based server placement scheme is employed to determine the optimal server-user ratio. The proposed scheme incentivizes ESs to serve EUs through all-pay auction, where EUs bid for server resources below market value. Then, the scheme determines the optimal ratio by maximizing the SP revenue.

\begin{figure}[tp]
\centering
\captionsetup{singlelinecheck = false, justification=justified}
\includegraphics[width=8cm]{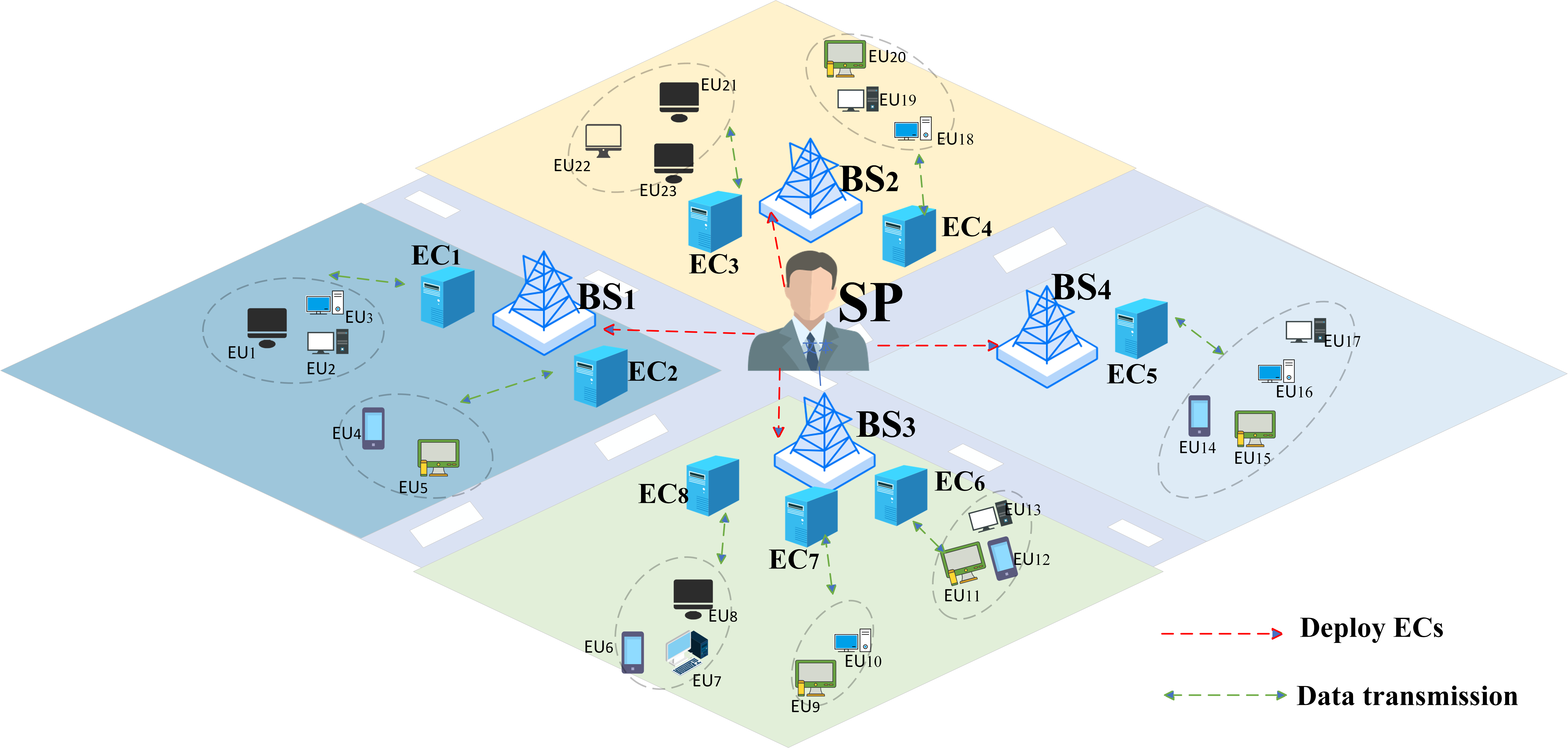}
\caption{System model.}
\label{sm}
\end{figure}

\section{All-pay Auction Framework-Based server placement Scheme}
As shown in Fig. \ref{sm}, we consider a multiserver-multiusers edge computing environment, consisting of a set of users $N$ = \{1, 2, ..., $|N|$\} and a set of MEC servers M = \{1, 2, ..., $|M|$\}. A server is responsibile for the  data offloading of some EUs in a region and these EUs form a set (refer to the \textbf{set allocation} algorithm in literature \cite{ref1}). EUs within the same set compete for server resources. Firstly, the EU evaluates the server resource according to the amount of offload data $q$ and the private valuation function $F(v)$. According to the rules of all-pay auction in literature \cite{ref1}, the EU further makes a equilibrium bid $b$ according to the valuation $v$. The EU with the largest bid in the same set gets the server service, and other EUs also need to pay their bids to the server. According to the number of EUs in different regions, the SP deploys the corresponding number of ESs to maximize its own revenue.



\begin{figure*}[!t]
\captionsetup{singlelinecheck = false, justification=justified, labelsep=period}
\subfigure[N = 100.]{
\begin{minipage}[t]{0.33\linewidth}
\centering
\includegraphics[width=2.56in]{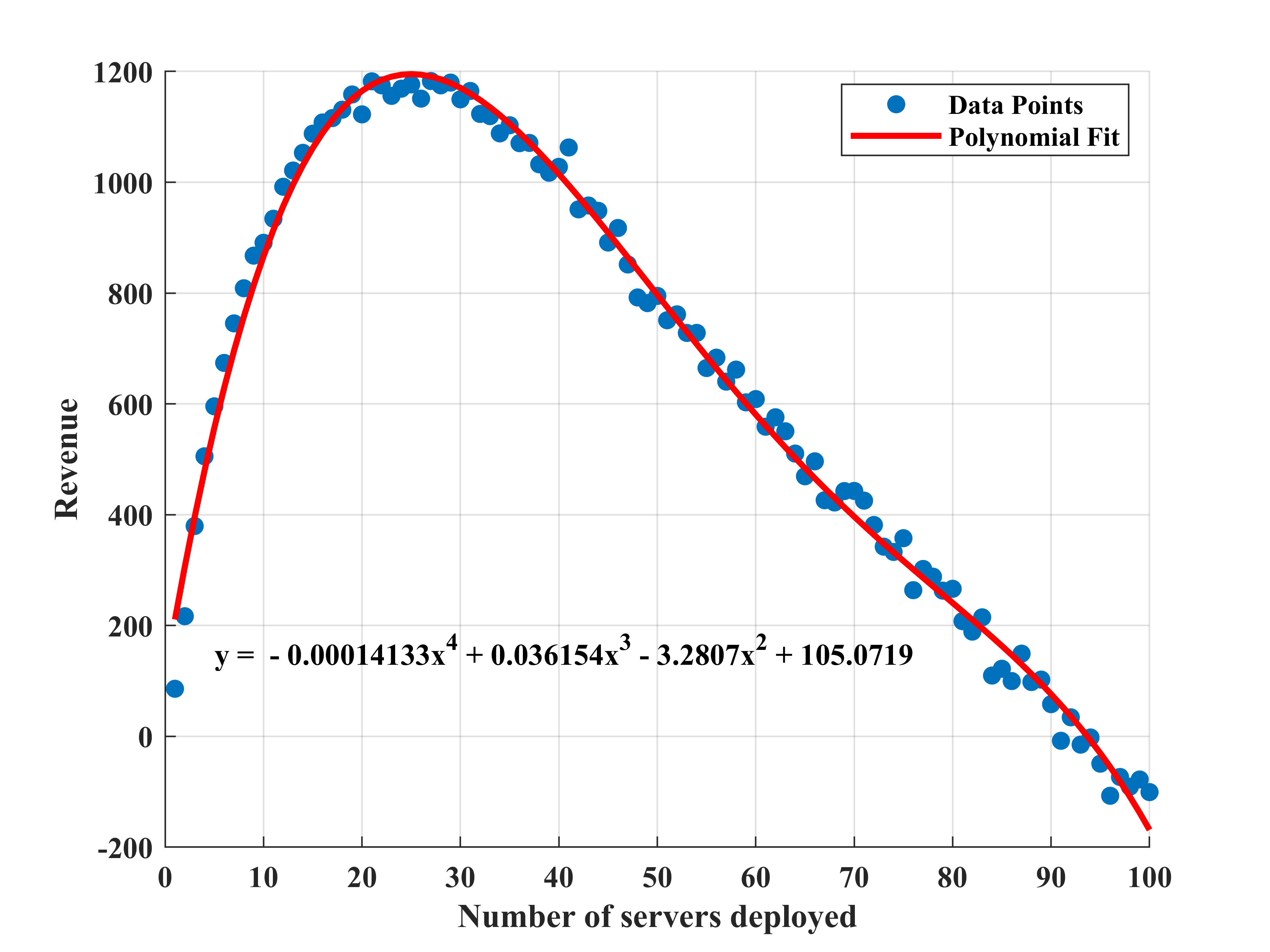}
\label{2.1}
\end{minipage}%
}%
\subfigure[N = 500.]{
\begin{minipage}[t]{0.33\linewidth}
\centering
\includegraphics[width=2.56in]{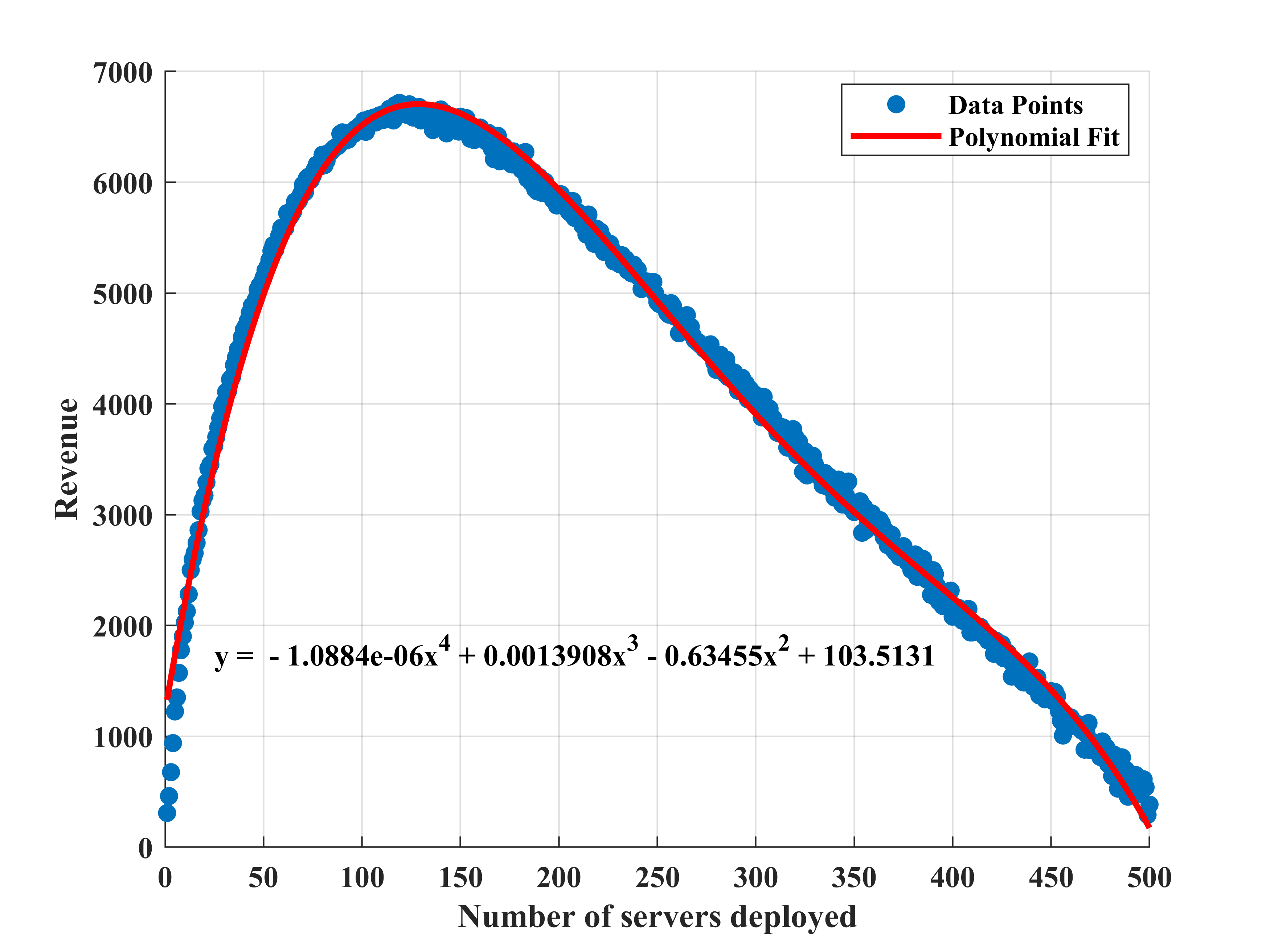}
\label{2.2}
\end{minipage}%
}%
\subfigure[N = 1000.]{
\begin{minipage}[t]{0.33\linewidth}
\centering
\includegraphics[width=2.56in]{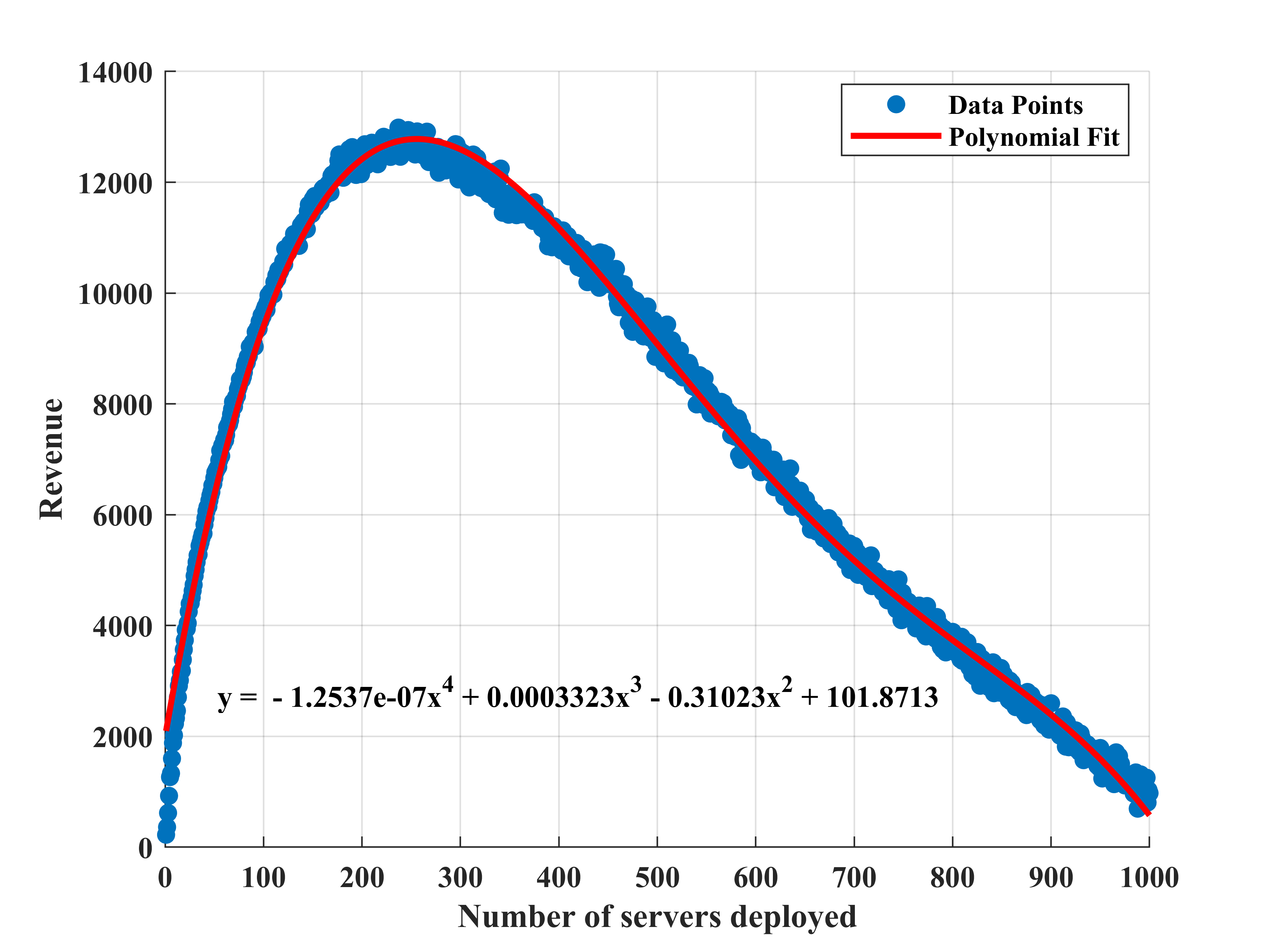}
\label{2.3}
\end{minipage}
}%
\centering
\caption{Effect of the number of servers on revenue.}
\label{22}
\end{figure*}
For the sake of expression simplicity, denote $F(v)$ of EU $n$ as the uniform distribution function over [0, $A$]: $F=\frac{v}{A}$, where $A$ represents the valuation ability of the EU. And we establish the relation between $A$ and the EU computation capacity to express this ability: $A=k\cdot lg F_t$, where $k > 0$ is a valuation coefficient.

Since the amount of offloaded data is larger, the valuation of the EU is larger. Denote that $Q$ is the maximum amount of data offloading allowed by the EU in each time slot. Assume that the relationship between valuation and offload is $v=A \cdot \sqrt{\frac{q}{Q}}$, where $q$ is the amount of offloaded data of EU.
Based on the all-pay auction \cite{ref1}, the \textbf{equilibrium bid} $b_e$  required by the EU to offload $q$ bit can be obtained:
$b_e = \int_0^v tdF^{n-1}(t)$, where $n$ is the number of EUs in the same set. Since the case of $n$ = 1 is not considered in the literature \cite{ref1}, we assume that when $n$ = 1, the $b_c$ = $e$. When $n$ = 1, if the ES chooses not to serve the EU, the ES still incurs a fixed cost $B$ and the EU's revenue is 0. If the ES chooses to serve the EU, the cost of the ES includes $B$ and the cost of processing the $q$ bit data, and the EU also generates revenue $u$. Then according to game theory \cite{GT}, the $b_c$ is obtained as follows.
\begin{equation}  
b_e=
\begin{cases}  
\int_0^v tdF^{n-1}(t), & n \geq 2 \\  
e, & n = 1 \\  
\end{cases} .  
\
\end{equation}
where $e$ $\in$ $[B+log_2q, u)$ and $u = v-e$.

The ES $j$ revenue consists of the fixed cost $B$, the cost of processing $q_w$ bit data and the bid of the EU $b_c$. The expression is as follows.

\begin{equation}
    U_{server}^j=\sum_{i=0}^{n_i}b_e^i - B - log_2q_w
\end{equation}
where $q_w$ is the amount of offloaded data corresponding to the winner and $n_i$ is the total number of EUs in the same set. However, if the number of ESs is too small, many EUs will not be served and SP has less benefit. On the contrary, there are many idle servers, causing a fixed cost and resulting in a decrease in revenue. Therefore, the SP should decide how many servers to deploy to maximize the total revenue and the revenue is expressed as follows.
\begin{equation}
    \sum W =\sum_{j=0}^M U_{server}^j
\end{equation}
\section{performance evaluation}
Fig. \ref{22} shows the overall revenue of SP deploying different number of ESs for different number of EUs. Denote that $N$ = 100, 500, 1000 and remaining parameters are uniformaly defined in Table \ref{tabel_1} \cite{ref1, ref2}. From Fig. \ref{22}, we can see that revenue shows a trend of first increasing and then decreasing. As the number of ESs increase, more and more EUs can be served in the same time period, and the total revenue increases. When it exceeds a certain amount, the total system revenue decreases. This is because, based on the \textbf{set allocation} algorithm, many ESs are idle, which cannot generate revenue while generating the fixed costs. And this is also the reason for the negative revenue in Fig. \ref{2.1}.
\begin{table}[!ht]
\begin{center}
\caption{parameter settings}
\label{tabel_1}
\begin{tabular}{|c|c|c|c|c|c|}  
\hline  
\textbf{Parameter}  & $q$  & $Q$ & $F_t$ & $k$ &  $B$ \\ \hline  
\textbf{Value} & [100, 500]KB & 500KB & \{0.01, 0.1, 1\}GHz  & 10 & 10 \\ \hline 
\end{tabular}
\end{center}
\end{table}

In addition, we fit the results in Fig. \ref{22}, and according to the fitted curve, the number of ESs deployment under the maximum benefit is obtained. That is, when the server-user ratio  is approximately  25\% , the revenue of SP is the largest.

\section{conclusion and future work}
This paper proposes a server placement scheme based on all-pay auction. By maximizing the system revenue, the optimal server-user ratio is obtained. Simulation results show that when the ratio is appropriately 25\%, the total system revenue is the largest. For our future work, we will implement
the proposed scheme into real-world equipment to acquire
practical results. Furthermore, we will explore more deeper
economic implications to investigate the long-term effects of
the proposed pricing scheme.

\end{document}